\documentclass[aps,prl,twocolumn,preprintnumbers,groupedaddress,nofootinbib,nobalancelastpage]{revtex4-2}

\usepackage{multirow}
\usepackage{amssymb}
\usepackage{enumerate}
\usepackage{amssymb}
\usepackage{amsmath}
\usepackage{tikz}
\usepackage{feynmf}
\usepackage{makecell}
\usepackage{slashed}
\usepackage{natbib}
\usepackage{graphicx}
\usepackage{changes,color}

\usepackage{mathrsfs,amssymb}  
\usepackage{tensor} 
\usepackage{cancel}
\usepackage[normalem]{ulem}
\usepackage{slashbox}
\usepackage{xcolor}

\definecolor{darkgreen}{rgb}{0.01, 0.75, 0.24}
\definecolor{darkorange}{rgb}{1.0, 0.55, 0.0}
\usepackage{pifont}
\newcommand{\beq}{\begin{equation}}
\newcommand{\eeq}{\end{equation}}
\newcommand\bea{\begin{eqnarray}}
\newcommand\eea{\end{eqnarray}}

\newcommand{\eq}[1]{Eq.~(\ref{#1})}

\newcommand{\ov}{\overline}
\renewcommand{\l}{\left}
\renewcommand{\r}{\right}
\newcommand{\la}{\langle}
\newcommand{\ra}{\rangle}

\begin{document}

\preprint{FTUV-24-02-07}
\preprint{IFIC-24-23}

\title{The minimal cosmological standard model}


\author{Gabriela Barenboim$^{1,2}$}
\email[]{gabriela.barenboim@uv.es}
\author{P. \ Ko$^3$} 
\email[]{pko@kias.re.kr}
\author{Wan-Il Park$^{4,1,2}$}
\email[]{wipark@jbnu.ac.kr}
\affiliation{
$^1$Instituto de Física Corpuscular, CSIC-Universitat de València, Paterna 46980, Spain\\
$^2$Departament de Física Teòrica, Universitat de València, Burjassot 46100, Spain	\\
$^3$Korea Institute for Advanced Study, Seoul 02455, Republic of Korea\\
$^4$Division of Science Education and Institute of Fusion Science, Jeonbuk National University, Jeonju 54896, Republic of Korea\\
}

\date{\today}

\begin{abstract}
We propose a novel minimal scenario which simultaneously addresses the following theoretical/cosmological/phenomenological puzzles: (i) the origin of scales, (ii) primordial inflation, (iii) matter-antimatter asymmetry, (iv) tiny neutrino masses, (v) dark matter, and (vi) the strong CP-problem.
Exact scale-symmetry was assumed.
A global $U(1)_{\rm PQ}$-symmetry was also assumed but only in the matter sector.
The novelty of the scenario is the introduction of explicit $U(1)_{\rm PQ}$-breaking terms with field-dependent coefficients in the gravity sector.
Such a term does not disturb the axion solution whereas naturally realizes an axi-majoron hybrid inflation which allows a natural realization of Affleck-Dine mechanism for generating Peccei-Quinn number asymmetry.
The asymmetry can be transferred to the visible sector via the right-handed neutrino portal through non-thermal decay and thermal processes, even without the presence of a CP-violating phase in the matter sector.
Dark matter and dark radiation are obtained by cold and hot components of axi-majorons, respectively.
\end{abstract}


\maketitle



\noindent
\textit{Introduction} - 
Our understanding of the cosmos is primarily governed by two widely accepted and tested theories, namely the Standard Model of Particle Physics (SM) and General Relativity (GR). Both theories revolve around their fundamental scales, the electroweak and Planck scales respectively, which are associated with discrete, constant mass parameters determined through empirical observations. However, the source of these parameters remains a mystery. Existing theories of high energy physics do not provide specific theoretical constraints, allowing for the determination of these scales beforehand. Therefore, it might be inherently logical to view scale-invariance as a key symmetry in the fabric of our universe. If this is the case, all scales, including the Planck scale, would arise from a spontaneous disruption of this symmetry.

In recent times, a variety of studies have delved into the dynamical generation of scales in a scale-invariant interpretation of gravity and expansions of the Standard Model (SM) \cite{Garcia-Bellido:2011kqb,Salvio:2014soa,Ferreira:2016vsc,Ferreira:2016wem}. 
One notable discovery is the dynamical generation of the Planck scale, even in the absence of Coleman-Weinberg (CW)-type symmetry-breaking, which is made possible due to the preservation of the scale current whose kernel eventually stabilizes to a constant \cite{Ferreira:2016wem}.

Lower energy scales, such as the electroweak scale, can be produced either in association with the field responsible for generating the Planck scale, or separately through the CW mechanism in the presence of matter fields distinct from the SM fields. Scale-invariance can still be maintained, even on a quantum level, as long as the renormalization scale (usually presented as an explicit dimensionful parameter) is substituted with a fundamental scalar field \cite{Ferreira:2016wem}. 

The maintenance of scale-invariance isolates the massless dilaton from other fields, effectively freeing the theory from fifth-force restrictions \cite{Blas:2011ac,Ferreira:2016kxi}. 
This suggests that a fully scale-invariant theory can be perfectly consistent  with low energy phenomenology.

Consequently, any theory attempting to elucidate the present condition of the Universe must be capable of incorporating a period of primordial inflation \footnote{A phenomenon which can
naturally arise in its slow roll formulation in the context of emergent gravity with scale-invariance \cite{Garcia-Bellido:2011kqb,Ferreira:2016vsc,Ferreira:2018qss} (see also Ref.~\cite{Kannike:2015apa}).}, 
address the generation of the matter-antimatter asymmetry, identify suitable dark matter candidate(s), accommodate neutrino masses and mixings as suggested by neutrino flavor 
oscillation data \cite{Gonzalez-Garcia:2007dlo}, and potentially provide a resolution to the strong CP-problem such as the invisible axion \cite{Peccei:1977hh,Peccei:1977ur,Kim:1979if,Shifman:1979if,Dine:1981rt,Zhitnitsky:1980tq}. 
These major conundrums in cosmology and phenomenology cannot be resolved solely within the SM and unequivocally call for a theory beyond the Standard Model(BSM).

In this work, we propose a minimal scale-invariant Peccei-Quinn(PQ)-symmetric extension of the SM.
Introducing a $U(1)_{\rm PQ}$-breaking term in the gravity sector only\footnote{A symmetry-breaking non-minimal gravitational interaction was considered in Refs.~\cite{Takahashi:2015waa,Hashimoto:2021xgu} but in different contexts.}, we show that our model can address the aforementioned issues altogether, providing a very simple unified framework for the unknown history of the universe from inflation to the big-bang nucleosynthesis.
Contrary to similar models in the literature (see for example Ref.~\cite{Ballesteros:2016xej}), our model is minimal and accommodates the dynamical generation of Planck scale via spontaneous breaking of scale-symmetry.

\vspace{1em}
\noindent
\textit{The model} - We assume that scale-invariance is a fundamental symmetry and preserved in both of the matter and the gravity sectors.
We also assume that $U(1)_{\rm PQ}$ is preserved but only in the matter sector, since gravity does not respect global symmetries.
The potential danger of strong non-perturbative breaking effects on a global symmetry may be retained negligible in the presence of Gauss-Bonnet term in the gravity sector \cite{Kallosh:1995hi}.
Based on these symmetries, we consider a scale-invariant Type-II two Higgs doublet extension of the SM \cite{Branco:2011iw}\footnote{KSVZ-type realization is also possible, and heavy extra quarks can contribute to dark matter relic density in this case.}, combined with the singlet majoron model \cite{Chikashige:1980ui} for the Type-I seesaw mechanism \cite{Minkowski:1977sc,Yanagida:1979as} (see Refs.~\cite{Volkas:1988cm,Clarke:2015bea} for earlier works with a similar setup). 
There are four scalar fields as the minimal set: a real scalar $\chi$ which is responsible for generating Planck scale via spontaneous breaking of scale symmetry, two Higgs doublets, and one complex singlet scalar which plays the role of the PQ-field, denoted here as $\Phi \equiv \phi e^{i \theta}/\sqrt{2}$ with its vacuum expectation value(VEV) $\phi_0 \equiv \la \phi \ra$ in the range of $\mathcal{O}(10^{9-12}) {\rm GeV}$.

Instead of considering all these fields explicitly, we deliver our argument with a simplified version having only the $\chi$ and $\Phi$ fields under the assumption that the vacuum expectation values of Higgs fields are negligible at high energy.
Paying attention to these fields only and their relevant interactions, we characterize our model by specifying the action of the gravity sector first as\footnote{The gravity part of Lagrangian can have additional scale-invariant curvature-terms such as ${\tilde R}^2, \ {\tilde R}_{\mu \nu} {\tilde R}^{\mu \nu}, \  {\tilde R}_{\mu \nu \rho \sigma} {\tilde R}^{\mu \nu \rho \sigma}$, and ${\tilde R}_{\mu \nu \rho \sigma} {}^*{\tilde R}^{\mu \nu \rho \sigma}$ where ${}^*{\tilde R}_{\mu \nu \rho \sigma} \equiv \epsilon_{\mu\nu\mu'\nu'}{\tilde R}\indices{^{\mu'}^{\nu'}_{\rho\sigma}}/2$.
They might be originated simply from two topological terms, ${\tilde R}_{\mu \nu \rho \sigma} {}^*{\tilde R}^{\mu \nu \rho \sigma}$ and ${}^*{\tilde R}_{\mu \nu \rho \sigma} {}^*{\tilde R}^{\mu \nu \rho \sigma}$.
Gauss-Bonnet term (i.e., the latter) may resolve the axion quality problem if the numerical coefficient is at least of order unity \cite{Kallosh:1995hi}.
We assume its presence, but omit it in the discussion, since it does not affect physics discussed in the body of this letter.
\label{foootnote:GB-term}},
\bea \label{eq:S-G}
S_{\rm G} &=& - \frac{1}{2} \int d^4 x \sqrt{-{\tilde g}} {\tilde R} \l[ \xi_\chi \chi^2 + 2 \xi_\phi |\Phi| ^2  \r.
\nonumber \\
&& \l. + \xi_+ \l( \Phi^2 + {\rm c.c.} \r) - i \xi_- \l( \Phi^2 - {\rm c.c.} \r) \r]
\nonumber \\
&\equiv& - \frac{1}{2} \int d^4 x \sqrt{-{\tilde g}} {\tilde R} \l[ \xi_\chi \chi^2 + \xi_\phi \phi^2 \l( 1 + \alpha \cos 2 \theta \r) \r] 
\eea
where ${\tilde R}$ and curvature-terms mentioned in the footnote \ref{foootnote:GB-term} are from the Jordan frame metric ${\tilde g}_{\mu \nu}$ with ${\tilde g} \equiv {\rm det} \l( {\tilde g}_{\mu \nu} \r)$, and the phase of $\Phi$ was redefined in the last expression.
The metric ${\tilde g}_{\mu \nu}$ is related to the metric in the Einstein frame, $g_{\mu \nu}$, by a conformal transformaton ${\tilde g}_{\mu \nu} = \Omega^{-2} g_{\mu \nu}$ where $\Omega$ is defined as
\beq
M_{\rm P}^2 \Omega^2 \equiv \xi_\chi \chi^2 + 2 \xi_\phi |\Phi| ^2  + \xi_+ \l( \Phi^2 + {\rm c.c.} \r) - i \xi_- \l( \Phi^2 - {\rm c.c.} \r)
\eeq
with $M_{\rm P} \simeq 2.4 \times 10^{18} {\rm GeV}$ being the reduced Planck scale.
In this case, one finds \cite{Wald:1984rg}
\beq
\Omega^{-2} {\tilde R} = R + 6 g^{\mu \nu} \l[ \l( \nabla_\mu \ln \Omega \r) \l( \nabla_\nu \ln \Omega \r) - \nabla_\mu \nabla_\nu \ln \Omega \r]
\eeq
where $R$ and $\nabla_\mu$ are the Ricci scalar and the covariant derivative in the Einstein frame, respectively.

All $\xi$s in \eq{eq:S-G} are assumed to be positive definite and real with $\alpha < 1$.
Especially, we assume that $\xi_\chi$ is a numerical constant whereas \footnote{In our simple ansaz of \eq{eq:xi-phi} for $\xi_\phi$, the field space where $\chi < 0$ can not provide a proper cosmology and low energy phenomenology since matter fields are pushed toward the origin.
The presence of such a case might not be a problem as long as it can not be selected in the sense of anthropic priciple \cite{Barrow:1986nmg}.
Alternatively, we could choose a different shape such as $\xi_0 \exp \l[ -c_\phi \l( \chi / \phi \r)^2 \r]$ for $\xi_\phi$.
Since we are not pursuing its origin in this work, we do not discuss the form of $\xi_\phi$ in \eq{eq:xi-phi} any further.}
\beq \label{eq:xi-phi}
\xi_\phi \equiv \xi_{\phi, 0} \exp\l( - c_\phi \chi/\phi \r)
\eeq
with $\xi_{\phi, 0}$ and $c_\phi$ being numerical constants, taking
\beq \label{eq:c-phi-range}
\sqrt{\xi_\chi} \phi_0/M_{\rm P} \ll c_\phi \ll \sqrt{\xi_\chi/\xi_{\phi,0}}
\eeq
to accommodate consistently low energy phenomenology and cosmology in a simple manner as discussed subsequently.
Such an exponential field-dependence might be from some non-perturbative physics, and has been used from time to time in the literature in different contexts such as Ref.~\cite{Zlatev:1998tr}.
In this work we do not pursue the origin of the form in \eq{eq:xi-phi}, but simply assume it as a possibility.

The other relevant parts of the model are the scalar potential and the seesaw sector interactions.
They are respectively given by
\bea \label{eq:V-scalar}
{\tilde V} &=& \frac{\lambda_\chi}{4} \chi^4 + \lambda_\phi |\Phi|^4 - \frac{1}{2} \lambda_{\chi \phi} \chi^2 |\Phi|^2
\\ \label{eq:V-fermion}
-\mathcal{L}_{\rm ss} &\supset& 
\frac{1}{2} y_N \Phi^* \ov{\nu_R^c} \nu_R + y_\nu \ov{\ell}_L {\tilde H}_2 \nu_R + {\rm h.c.}
\eea
where $\nu_R$ is the right-handed neutrino(RHN) field, $\ell_L$ is the SM left-handed lepton doublet, ${\tilde H}_2$ is the conjugate of the up-type Higgs doublet $H_2$, and the flavor indices were omitted.
The  $\lambda$'s in \eq{eq:V-scalar} are assumed to be positive definite real free parameters.
Note that, thanks to the Yukawa interaction $y_N$, the spontaneous breaking of $U(1)_{\rm PQ}$ is responsible for Majorana masses of RHN fields.
Because of that, we will call the associated pseudo-Goldstone boson as \textit{axi-majoron}\footnote{See Ref.~\cite{Shin:1987xc} for earlier work.}.

The reduced Planck scale can be determined mainly by $\chi_0$ the vacuum expectation value(VEV)s of $\chi$ as $M_{\rm P} \simeq \xi_\chi^{1/2} \chi_0$ via a spontaneous breaking of scale-invariance thanks to the presence of a constant kernel of the scale current as discussed in Ref.~\cite{Ferreira:2016wem}.
Applying the idea of Ref.~\cite{Ferreira:2016wem} to our scenario, we notice that, if $\xi_\chi$ is sufficiently smaller than unity, it is possible to have Planck scale nearly fixed while inflation is driven by $\Phi$ even with trans-Planckian initial condition for inflation \cite{Barenboim:2024xxa}. 
In such a case, the situation becomes essentially the same as the case of non-minimal gravitational couplings of $\Phi$ with the (nearly) fixed Planck scale. 
We take this approximation in the subsequent discussion.

The novel aspects of our scenario are the presence of the symmetry-breaking non-minimal gravitational interaction, $\alpha$-term, and the $\phi$-dependence of $\xi_\phi$ in \eq{eq:xi-phi}.
When $c_\phi \chi/\phi \ll 1$, the $\alpha$-term provides a temporal mass to axi-majoron, and has following multiple impacts:
(i) realization of a large-field axi-majoron hybrid inflation, generating angular motion of the inflaton, the PQ-field, during and at the end of inflation,
(ii) removal of domain-wall problem due to the angular motion.
On the other hand, when $c_\phi \chi/\phi \ggg 1$ is satisfied at low energy, the $\xi_\phi$-terms are essentially turned off and the axion solution to the strong CP-problem remains untouched.
Note that, if $\xi_\phi$ were a pure numerical constant, unless $\xi_\phi \alpha$ is smaller than unity by many orders of magnitude, the axion solution can not work due to a large extra contribution to the axion mass that is caused by radiative corrections induced by graviton-mediated interactions \cite{Hill:2020oaj} (see Ref.~\cite{Barenboim:2024xxa}).
The field-dependence  of $\xi_\phi$ remedies such a danger, retaining the axion solution still valid. 
Iso-curvature perturbations are also suppressed because there is no instability causing a growth of the modes while the speed of the inflaton increases toward the end of inflation \cite{Barenboim:2024xxa}.

\vspace{1em}
\noindent
\textit{Inflation} - A large-field primordial inflation is realized along the PQ-field, but it is of a type of hybrid inflation involving the axi-majoron field too.
Depending on the magnitude of $\alpha$, angular motion in the complex field space of $\Phi$ can appear during and/or at the end of inflation.
For $\alpha \lll 1$ inflation is essentially the same as the SM Higgs-inflation \cite{Bezrukov:2007ep} with the nearly same predictions for inflationary observables. 
The expansion rate during inflation is given by $H_I \simeq \lambda_\phi^{1/2} M_{\rm P} / 2 \sqrt{3} \xi_{\phi,0}$ and inflationary observables matching observations such as Planck mission \cite{Planck:2018jri} is obtained if
\beq \label{eq:lambda-xi-relation}
\lambda_\phi \approx 6 \times 10^{-10} \xi_{\phi,0}^2
\eeq
We will be interested in $\xi_{\phi,0} = \mathcal{O}(1-10^2)$ for post-inflation cosmology.
The unitary bound can be lower than the Planck scale \cite{Burgess:2009ea,Burgess:2010zq,Barbon:2009ya}, but it is still well above the scales involved in our scenario.
 \begin{figure}[ht] 
\begin{center}
\includegraphics[width=0.48\textwidth]{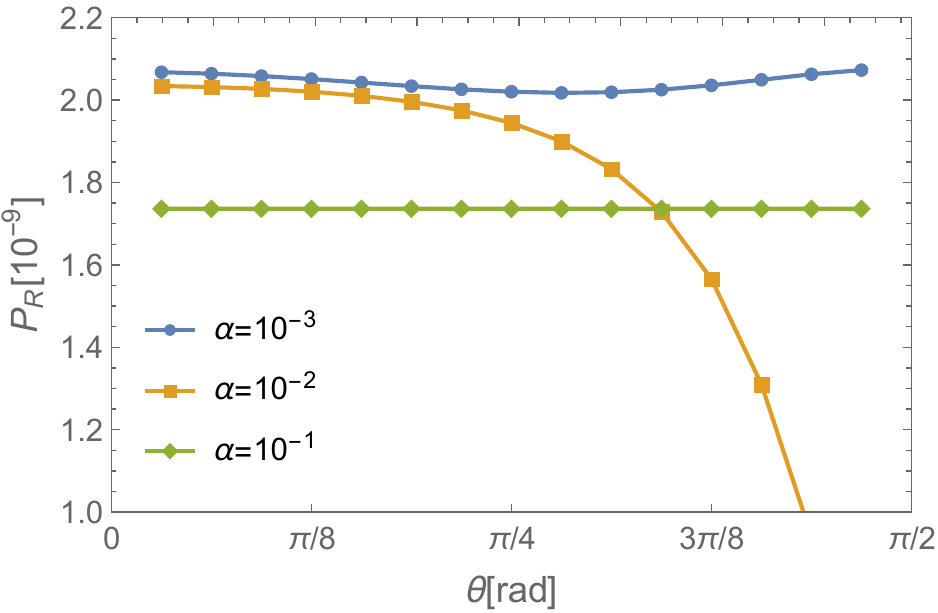}
\hspace{1em}
\includegraphics[width=0.48\textwidth]{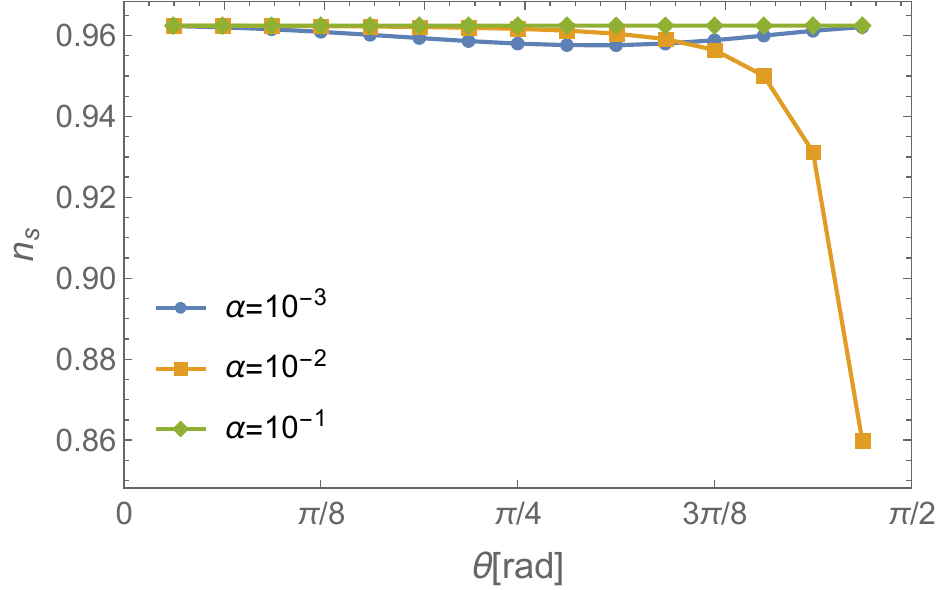}
\caption{
The amplitude of the power spectrum $P_R$(top) and the spectral index $n_s$(bottom) of density perturbations as functions of $\theta_{\rm ini}$ for various values of $\alpha$ with $\xi_{\phi,0}=1$, $\lambda_\phi=6 \times 10^{-10}$, $\phi_{\rm ini}=10 M_{\rm P}$.
}
\label{fig:para-dependence}
\end{center}
\end{figure}
As shown in Fig.~\ref{fig:para-dependence}, for $\alpha \xi_{\phi,0} \ll \mathcal{O}(10^{-2})$, inflationary observables are barely affected.
If $\alpha \xi_{\phi,0} \gtrsim \mathcal{O}(10^{-2})$, a dependence on the initial angular position $\theta_{\rm ini}$ and distance $\phi_{\rm ini}$ appear.
As long as $\theta_{\rm ini}$ is not very close to $\pm \pi/2$ (i.e., the potential maxima along the angular direction) and $\phi_{\rm ini}$ is large enough to accommodate enough amount of $e$-folds, the motion of the inflaton in the 2-D field space is smooth enough without any rapid turn of the trajectory.
Also, the effective mass of the fluctuations orthogonal to the inflaton direction has a positive definite sign.
As a result, there is no sizable impact of a two-dimensional curvilinear trajectory of the inflaton on the evolutions of adiabatic and iso-curvature perturbations \cite{Barenboim:2024xxa}.

The energy of inflaton is initially stored dominantly in the radial zero mode of $\Phi$ although some fraction is in the angular zero mode.
The preheating effect right after inflation may develop quite large fluctuations along both radial and angular directions of $\Phi$, leading to symmetry restoration \cite{Tkachev:1998dc}.
Hence, there might be a worry of the domain wall problem after the breaking of the restored symmetry.
However, it should be noted that the symmetry-breaking is not exactly same as the case of a field stuck at the origin due to thermal effect.
The dynamical background mode can still exist in addition to the sizable fluctuations as long as it is not thermalized and decays only after the breaking of the restored symmetry.
In particular, if $\Phi$ has a non-zero angular momentum sufficiently larger than the contribution from quantum fluctuations, as long as the angular momentum is conserved after inflation as in our scenario, the background motion of $\Phi$ does not pass through the origin of the potential.
The condition for a sizable angular momentum can be expressed as
\beq \label{eq:cond-for-no-domain-wall}
\l| \frac{\dot{\theta}_e}{H_e} \r| \gg \frac{H_e}{2 \pi M_{\rm P}}
\eeq
where $\dot{\theta}_e$ and $H_e$ are the phase velocity of $\Phi$ and the expansion rate  at the end of inflation, respectively.
For a sufficiently small $\alpha \xi_{\phi, 0}$ satisfying the slow-roll condition along the phase direction, it is straightforward to find that the condition in \eq{eq:cond-for-no-domain-wall} corresponds to 
\beq \label{eq:cond-for-alpha}
%
\frac{3 \times 10^{-7}}{\sin \l( 2 \theta_e \r)} \ll \alpha \xi_{\phi, 0} \lesssim \mathcal{O}(10^{-2})
\eeq
where \eq{eq:lambda-xi-relation} and $H_e \approx H_I$ were used for the lower bound.
Hence, there exists $\alpha \xi_{\phi, 0}$ satisfying \eq{eq:cond-for-alpha} as long as $\theta_e \gg \mathcal{O}(10^{-5})$ which is much larger than the size of quantum fluctuations during inflation.
Also, the non-equilibrium condition before the phase transition of the restored symmetry can be satisfied as long as the interactions of the inflaton with right-handed neutrinos are sufficiently small as will be discussed subsequently.
In this circumstance, for $\alpha$ satisfying \eq{eq:cond-for-alpha}, the vacuum positions in the phase space of $\Phi$ after symmetry-breaking are expected not to be  random but to have a biased distribution centered at the position associated with the background mode.
It might be possible for an infinite cosmic string to be formed \cite{Tkachev:1998dc}, and then domain walls would appear at the epoch of the QCD phase transition.
However, as discussed in Ref.~\cite{Coulson:1995nv}, in case of the biased vacuum distribution the string with walls may decay, removing the domain wall problem \footnote{
Taken into account of the effect of topological defects, the symmetry-breaking scale of $U(1)_{\rm PQ}$ may have to be adjusted to match observations in regard of axion dark matter.
Interestingly, if realized, the decay of the string-domain wall system may leave some observable impacts such as a gravitational wave background relevant for NANOGrav discovery \cite{NANOGrav:2023gor}.
We will touch this issue in other work.}. 

In the absence of a sizable coupling to the SM Higgs field, that is what we assume, the standard thermal background after inflation can be established only if the inflaton can decay to RHNs too in addition to axi-majorons. 
For a RHN mass-eigenstate $N_i$, it requires
\beq \label{eq:cond-for-RHN-ch}
y_{N_i} < \sqrt{\lambda_\phi } \ (\textrm{or} \ \sqrt{3 \lambda_\phi /2})
\eeq
with the phase space factor in the decay rate ignored.
For $y_{N_i}$ satisfying \eq{eq:cond-for-RHN-ch}, even if the fluctuations of $\Phi$ during preheating is ignored,  the symmetry-restoration can take place for a range of $\xi_{\phi, 0}$ \cite{Barenboim:2024xxa}, but again a sizable angular motion of $\Phi$ can save us from the domain wall problem as long as $\Phi$ does not decay before the breaking of the restored symmetry while it is still out of equilibrium.

For the largest $y_{N_i}$ denoted as $y_N$, if $y_N \ll \sqrt{\lambda_\phi}$, the decay rate of the dominant radial mode of $\Phi$ to right-handed neutrinos is approximately given by
\beq
\Gamma_\phi \simeq \frac{y_N^2}{32 \pi} m_\phi
\eeq
where $m_\phi$ is the mass of $\phi$.
Meanwhile, for $\chi_0$ fixed, the low energy potential of $\Phi$ in the sub-Planckian field space can be cast in the form, ${\tilde V} = \lambda_\phi \l( \phi^2 - \phi_0^2 \r)^2 / 4$. 
Taking into account of the field fluctuation ($\delta \phi$), the potential of the homogeneous mode ($\ov{\phi}$), $\ov{V} \equiv \langle {\tilde V} \rangle$, is found to be 
\beq
\ov{V} = \frac{\lambda_\phi}{4} \l[ \l( \ov{\phi}^2 - \phi_0^2 + 3 \ov{\l( \delta \phi \r)^2} \r)^2 + 4 \phi_0^2 \ov{\l( \delta \phi \r)^2} - 8 \ov{\l( \delta \phi \r)^4} \r]
\eeq
where $\ov{\l( \delta \phi \r)^2} \equiv \langle \l( \delta \phi \r)^2 \rangle$.
As the field fluctuations are red-shifted, satisfying $\ov{\l( \delta \phi \r)^2} < \ov{\l( \delta \phi \r)^2_{\rm c}} \equiv \phi_0^2/3$, the curvature of the potential at the origin becomes negative from positive, and vacua at $\ov{\phi} \neq 0$ appear.
For $\ov{\l( \delta \phi \r)^2} \approx \ov{\phi}_{\rm osc}^2$ with $\ov{\phi}_{\rm osc}$ being the oscillation amplitude of the background mode \cite{Tkachev:1998dc}, $\ov{\phi}$ does not get over the origin if $\ov{\phi}_{\rm osc} < \ov{\phi}_\times \equiv \sqrt{2/7} \phi_0$.
The expansion rate at this epoch is 
\beq
H_\times \sim \frac{\ov{V}(\ov{\phi}_\times)^{1/2}}{\sqrt{3} M_{\rm P}} \approx \frac{5 \lambda_\phi^{1/2}}{14 \sqrt{3}} \l( \frac{\phi_0^2}{M_{\rm P}} \r)
\eeq
Then, requiring $\Gamma_\phi(\ov{\phi}_\times) / H_\times < 1$ with $m_\phi(\ov{\phi}_\times) = \sqrt{5 \lambda_\phi / 7} \phi_0$ as the condition for non-equilibrium of the background mode with \eq{eq:lambda-xi-relation} satisfied, one finds a constraint on the non-minimal coupling $\xi_{\phi,0}$ as
\beq
\xi_{\phi, 0} \lesssim 41 \l( \frac{0.1}{b} \r)^{1/2} \l( \frac{\phi_0}{10^{10} {\rm GeV}} \r)^{1/2}
\eeq
where $b \equiv y_N^2/\lambda_\phi$.
Hence, for $b \lesssim \mathcal{O}(0.1)$ and $\phi_0 = \mathcal{O}(10^{9-12}) {\rm GeV}$, the non-equilibrium condition for a biased symmetry-breaking can be satisfied for $\xi_{\phi, 0} = \mathcal{O}(1 - 10^2)$ which is of our interest.

As discussed in the previous paragraphs, the inflaton may have to decay during the matter-domination(MD) era while it is in the broken phase, producing axi-majorons and RHNs.
The expansion rate at the epoch of inflaton's decay is then expected to be $H_* \approx \l( \sqrt{2}/ 24 \pi \r) \lambda_\phi^{3/2} \phi_0$ \cite{Barenboim:2024xxa}.
The standard thermal bath can be recovered if at least one of the 
RHN mass-eigenstates has a  lifetime long enough and decays when it dominates the universe.
The decay rate of $N_i$ having mass $m_{N_i}$ much larger than the electroweak scale is given by
\beq \label{eq:GammaN-heavy}
\Gamma_{N_i} \simeq \frac{m_{\nu_i}}{4\pi} \l( \frac{m_{N_i}}{v_u} \r)^2
\eeq
where we used the seesaw relation for the mass of the left-handed neutrinos (LHNs)\footnote{For simplicity we assumed that $y_{\nu_i}$ is nearly flavor-diagonal.}, $m_{\nu_i} \approx y_{\nu_i}^2 v_u^2/ 2 m_{N_i}$ with $v_u/\sqrt{2}$ being the VEV of the neutral component of the up-type Higgs doublet.
For heavy RHNs we may take $m_{\nu_i} \sim m_\nu \equiv 0.05 {\rm eV}$ \cite{Gonzalez-Garcia:2007dlo}. 
For $\phi_0 = \mathcal{O}(10^{9-12}) {\rm GeV}$, unless $y_{N_i}^2/\lambda_\phi$ is smaller than unity by several orders of magnitude, those states are expected to decay well before the EWPT while they are still relativistic and never dominate the universe.  
On the other hand, $m_{\nu_1}$ can be much smaller than $m_\nu$ by many orders of magnitude.
In such a case, the lightest RHN $N_1$ can start dominating when the expansion rate becomes $H_{\rm eq} \approx 2 \sqrt{2} {\rm B}_1^3 H_*$ with ${\rm B}_1 \approx y_{N1}^2/2 \lambda_\phi$ being the branching fraction of $\phi$ to $N_1$ \cite{Barenboim:2024xxa}.
Eventually, the decay of $N_1$ recovers the standard thermal bath.

\begin{figure}[ht] 
\begin{center}
\includegraphics[width=0.48\textwidth]{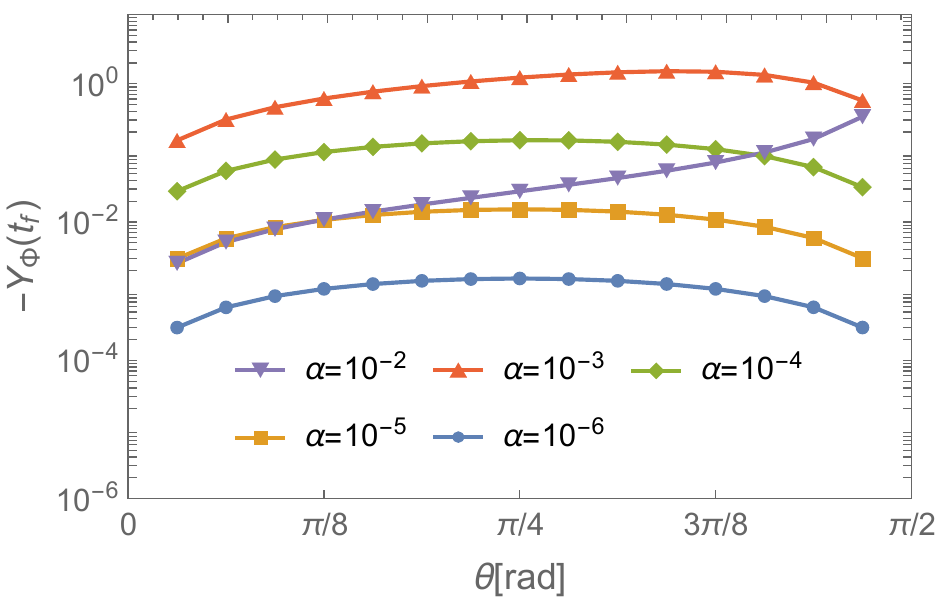}
\caption{
The yield of the PQ-number asymmetry of $\Phi$, $Y_\Phi(t_f)$, at $t_f = 10^4 H_{\rm ref}^{-1}$ as functions of $\theta_{\rm ini}$ for various $\alpha$ with the same parameter set as the one in Fig.~\ref{fig:para-dependence}.
$H_{\rm ref}$ is the expansion rate at the beginning of the field-evolution at ($\phi_{\rm ini}$, $\theta_{\rm ini}$).
Inflation ends at $t_e = \mathcal{O}(100) H_{\rm ref}^{-1}$ depending on initial position of the inflaton and $\alpha$.
}
\label{fig:YPQ-vs-theta-ini}
\end{center}
\end{figure}

\vspace{1em}
\noindent
\textit{Leptogenesis} - As the key feature of our scenario, thanks to the symmetry-breaking $\alpha$-term in \eq{eq:S-G}, a sizable amount of angular kick can be generated at the end of inflation as long as $\theta_{\rm ini}$ is not very close to zero and the mass scale along the angular direction is somewhat smaller but not extremely smaller than the expansion rate during inflation.
The angular motion is conserved after inflation, since the potential preserves $U(1)_{\rm PQ}$-symmetry.
This is nothing but the  Affleck-Dine(AD) mechanism \cite{Affleck:1984fy}, generating a PQ-number asymmetry.
Fig.~\ref{fig:YPQ-vs-theta-ini} as one of the key points of our scenario shows numerical estimations of the PQ-number asymmetry $Y_\Phi \approx Y_{\Phi, e}$ well after inflation for various 
values of $\alpha$ as functions of $\theta_{\rm ini}$.
The very different behavior in the case of $\alpha=10^{-2}$ is due to the fact that, when $\alpha \xi_{\phi,0}$ is somewhat large, the angular motion during inflation is sizable and makes the angular position at the end of inflation closer to $\theta=0$.

The asymmetry is converted to the lepton number asymmetry of the visible sector through the seesaw sector.
Depending on the nature of RHNs' decay, the conversion process can be like the usual Affleck-Dine mechanism (i.e., transfer of asymmetry through non-thermal decays) or spontaneous baryogenesis \cite{Cohen:1987vi}.
The heavy RHN states, $N_{2,3}$, are thermalized with the SM particles after their decays, but $N_1$ is never thermalized, since the requirement of recovering the standard thermal bath at late time without too much dark radiation constrains $y_{\nu_1}$ such that $m_{\nu_1} / m_\nu \lesssim \mathcal{O}(10^{-4} - 10^{-3})$.
There is a contribution of $N_{2,3}$ to the baryon number asymmetry via spontaneous leptogenesis \cite{Chun:2023eqc} (see also Refs.~\cite{Chiba:2003vp,Co:2019wyp}).
It can be large enough to explain the observed baryon number asymmetry by itself.
However, for a given set of parameters, it is overwhelmed by the contribution from the transfer of the helicity asymmetry of $N_1$ to the visible sector via non-thermal decays as in a usual Affleck-Dine mechanism \cite{Barenboim:2024xxa}.

The late time baryon number asymmetry in terms of yield can be expressed as $Y_{B, {\rm AD}} = \l(12/37 \r) \times \beta Y_{1, \rm d}$ where $\beta$ and $Y_{1, \rm d}$ are the speed and the yield of the helicity asymmetry of $N_1$ at its decay, respectively. 
The presence of $\beta$ relative to the case of the conventional AD-mechanism is due to the helicity mixing effect in the decay of RHNs through the Yukawa interaction $y_\nu$.
In the case inflaton decays in the broken phase, the asymmetry is then found as \cite{Barenboim:2024xxa}
\bea \label{eq:YB-AD-N1-MD}
Y_{B, {\rm AD}} 
&\simeq& 8.4 \times 10^{-11} \times \l( \frac{0.1}{{\rm B}_1} \r)^{4/3} \l( \frac{\xi_{\phi,0}}{10} \r)^{19/6}
\nonumber \\
&& \times \l( \frac{10^4 m_{\nu_1}}{m_\nu} \r)^{7/6} \l( \frac{\phi_0^{\rm ref}}{\phi_0} \r) \l( \frac{Y_{\Phi, e}}{10^{-2}} \r) \ .
\eea
From \eq{eq:YB-AD-N1-MD} with Fig.~\ref{fig:YPQ-vs-theta-ini}, one can see that the observed baryon number asymmetry can be explained for a wide range of $\alpha$, depending mainly on $B_1, \xi_{\phi, 0}$, and $\phi_0$.

\vspace{1em}
\noindent
\textit{Dark matter and dark radiation} - Dark matter in our scenario is only from cold axi-majorons produced via the misalignment mechanism \footnote{If there appears an infinite cosmic string at the breaking of the restored symmetry after inflation, contributions from decays of strings and domain walls would have to be taken into account.
Realization of such a possibility may depend on the angular momentum of the inflaton at the end of inflation, and requires a detailed numerical study.
It will be dealt with in a future work.}.
Hence, for $\theta_{\rm mis} = \mathcal{O}(1)$ as the misalignment angle, the symmetry-breaking scale of $U(1)_{\rm PQ}$ is required to be $\phi_0 = \mathcal{O}(10^{12}) {\rm GeV}$ \cite{Marsh:2015xka}.
Also, a sizable amount of dark radiation can be obtained naturally from the hot axi-majorons produced in the decay of the inflaton.
Its fractional energy density after the decay of $N_1$ is determined by the duration of the MD-era.
In terms of the conventional $\Delta N_{\rm eff}$ counting extra neutrino species, it can be given by
\bea
\Delta N_{\rm eff} 
&\simeq& 0.47 \times \l( \frac{10 \Gamma_{N_1}}{H_{\rm eq}} \r)^{\frac{2}{3}}
\nonumber \\
&\simeq& 0.3 \times \l[ \l( \frac{0.1}{{\rm B}_1^2 \xi_{\phi,0}} \r) \l( \frac{10^4 m_{\nu_1}}{m_\nu} \r) \frac{\phi_0}{\phi_0^{\rm ref}} \r]^{\frac{2}{3}} \ .
\eea
The extra radiation contributed by hot axi-majorons may help to ameliorate the tension in the observations of the expansion rate of the present universe \cite{Vagnozzi:2019ezj,DiValentino:2021izs}.
Note that the presence of an early MD-era causes a shift of the spectral index of density perturbations($n_s$) to a value smaller than the case of the standard cosmology.
Hence, in principle, $\Delta N_{\rm eff}$ can be used for a consistency check of our scenario with $n_s$ although it depends on the sensitivity of experiments.
Also, an early MD-era causes specific spectral changes of the nearly scale-invariant inflationary gravitational waves which contain information of $\Gamma_{N_1}$ and $H_{\rm eq}$.
Combination of $\Delta N_{\rm eff}$ and the spectral information of GWs with inflationary observables would allow extracting information of $m_{\nu_1}$ and $m_{N_1}$.
This may be a unique way of probing those quantities.

\vspace{1em}
\noindent
\textit{Conclusions} - 
In this work we proposed a full scale-invariant minimal scenario beyond the standard model with $U(1)_{\rm PQ}$-symmetry imposed only on the matter sector. 
We introduced \textit{$U(1)_{\rm PQ}$-breaking terms with field-dependent coefficients only in the gravity sector,} and showed that the model can simultaneously address the following theoretical, cosmological, and phenomenological puzzles: (i) the origin of scales, (ii) a primordial inflation, (iii) matter-antimatter asymmetry, (iv) tiny neutrino masses, (v) dark matter, (vi) strong CP-problem.
Specially, the symmetry-breaking non-minimal gravitational interaction allows a realization of an axi-majoron hybrid inflation along the PQ-field direction, which naturally induces a generation of a large amount of PQ-number asymmetry at the end of inflation.
The asymmetry is eventually converted to the baryon number asymmetry of the visible sector through the seesaw sector.
The standard thermal bath is to be recovered by decays of the lightest right-handed neutrinos.
Hot axi-majorons produced in the decay of the inflaton can play the role of dark radiation, and may ameliorate the Hubble tension \cite{Vagnozzi:2019ezj,DiValentino:2021izs}. 
Dark matter is made of purely cold axi-majorons produced through the conventional misalignment mechanism only.

Axion-solution is still valid because the symmetry-breaking terms are turned off at low energy in the late universe, thanks to the field-dependent nature of couplings which we assumed in this work.
Axion-quality problem may be absent in the presence of Gauss-Bonnet term in the gravity sector.
It is expected that the model is free from the domain wall problem as long as the PQ-field has a sizable angular motion that leads to a biased breaking of $U(1)_{\rm PQ}$-symmetry even though the symmetry may be restored due to preheating effects after inflation.
Iso-curvature perturbations of axi-majoron are naturally suppressed since there is no instability causing a growth of modes while the speed of inflation increases towards the end of inflation.
Anomalous-interactions of axi-majoron with gauge-field do not cause a large enhancement of gauge-field perturbations, since the speed of the axi-majoron during inflation is kept smaller than the expansion rate by several orders of magnitude.

Our scenario as a minimal extension of the standard model provides a very simple unified framework for the unknown history of the universe from inflation to big-bang nucleosynthesis, thanks to a full scale-symmetry and a $U(1)_{\rm PQ}$-breaking non-minimal gravitational interaction.
This model predicts a presence of a short early matter-domination era, which would leave a characteristic fingerprint on inflationary gravitational waves(GWs). 
In particular, once detected, GWs in combination with dark radiation and inflationary observables may provide information of the masses of the lightest left-handed and the lightest right-handed neutrinos.

\medskip


\medskip


\section*{Acknowledgments}

\noindent
This work was supported by the National Research Foundation of Korea grants by the Korea government: 2017R1D1A1B06035959 (W.I.P.),  2022R1A4A5030362 (W.I.P.), 
2019R1A2C3005009 (P.K.) and KIAS Individual Grants under Grant No. PG021403 (P.K.).
It was also supported by the Spanish grants PID2020-113775GB-I00 (AEI/10.13039/501100011033) and CIPROM/2021/054 (Generalitat Valenciana) (G.B and W.I.P.)
and  by the European ITN project HIDDeN (H2020-MSCA-ITN-2019/860881-HIDDeN) (G.B.).

\vspace{1em}
\bibliographystyle{apsrev4-2}
\bibliography{mcsm-ref}

\end{document}